\documentclass[12pt]{article}
\usepackage{graphicx}
\usepackage{lscape}
\usepackage{citesort}

\begin{document}

\def\ds{\displaystyle}
\def\beq{\begin{equation}}
\def\eeq{\end{equation}}
\def\bea{\begin{eqnarray}}
\def\eea{\end{eqnarray}}
\def\beeq{\begin{eqnarray}}
\def\eeeq{\end{eqnarray}}
\def\ve{\vert}
\def\vel{\left|}
\def\ver{\right|}
\def\nnb{\nonumber}
\def\ga{\left(}
\def\dr{\right)}
\def\aga{\left\{}
\def\adr{\right\}}
\def\lla{\left<}
\def\rra{\right>}
\def\rar{\rightarrow}
\def\nnb{\nonumber}
\def\la{\langle}
\def\ra{\rangle}
\def\ba{\begin{array}}
\def\ea{\end{array}}
\def\tr{\mbox{Tr}}
\def\ssp{{\Sigma^{*+}}}
\def\sso{{\Sigma^{*0}}}
\def\ssm{{\Sigma^{*-}}}
\def\xis0{{\Xi^{*0}}}
\def\xism{{\Xi^{*-}}}
\def\qs{\la \bar s s \ra}
\def\qu{\la \bar u u \ra}
\def\qd{\la \bar d d \ra}
\def\qq{\la \bar q q \ra}
\def\gGgG{\la g^2 G^2 \ra}
\def\q{\gamma_5 \not\!q}
\def\x{\gamma_5 \not\!x}
\def\g5{\gamma_5}
\def\sb{S_Q^{cf}}
\def\sd{S_d^{be}}
\def\su{S_u^{ad}}
\def\ss{S_s^{??}}
\def\sbp{{S}_Q^{'cf}}
\def\sdp{{S}_d^{'be}}
\def\sup{{S}_u^{'ad}}
\def\ssp{{S}_s^{'??}}
\def\sig{\sigma_{\mu \nu} \gamma_5 p^\mu q^\nu}
\def\fo{f_0(\frac{s_0}{M^2})}
\def\ffi{f_1(\frac{s_0}{M^2})}
\def\fii{f_2(\frac{s_0}{M^2})}
\def\O{{\cal O}}
\def\sl{{\Sigma^0 \Lambda}}
\def\es{\!\!\! &=& \!\!\!}
\def\ap{\!\!\! &\approx& \!\!\!}
\def\ar{&+& \!\!\!}
\def\ek{&-& \!\!\!}
\def\kek{\!\!\!&-& \!\!\!}
\def\cp{&\times& \!\!\!}
\def\se{\!\!\! &\simeq& \!\!\!}
\def\eqv{&\equiv& \!\!\!}
\def\kpm{&\pm& \!\!\!}
\def\kmp{&\mp& \!\!\!}

\def\pp{p^\prime}
\def\pps{p^{\prime 2}}

\def\sp{s^\prime}
\def\sps{s^{\prime 2}}


\def\simlt{\stackrel{<}{{}_\sim}}
\def\simgt{\stackrel{>}{{}_\sim}}


\title{
         {\Large
                 {\bf
Semileptonic decays of pseudoscalar mesons to scalar $f_0$ meson
                 }
         }
      }

\author{\vspace{1cm}\\
{\small T. M. Aliev \thanks
{e-mail: taliev@metu.edu.tr}~\footnote{permanent address:Institute
of Physics,Baku,Azerbaijan}\,\,,
M. Savc{\i} \thanks
{e-mail: savci@metu.edu.tr}} \\
{\small Physics Department, Middle East Technical University,
06531 Ankara, Turkey} }

\date{}

\begin{titlepage}
\maketitle
\thispagestyle{empty}

\begin{abstract}
The transition form factors of $D_s \rar f_0 \ell \nu$,
$D \rar f_0 \ell \nu$ and $B_u \rar f_0 \ell \nu$ decays are calculated in
3--point QCD sum rule method, assuming that $f_0$ is a quark--antiquark
state with a mixture of strange and light quarks. The branching ratios of
these decays are calculated in terms of the mixing angle.
\end{abstract}

\end{titlepage}

\section{Introduction}

The inner structure of the scalar mesons in terms of quarks is still an open
question in particle physics and it is the subject of an intense and
continuous theoretical and experimental investigations for establishing
their their nature (for a review, see \cite{R8201}). There are numerous
scenarios for the classification of the scalar mesons. The established
$0^{++}$ mesons are divided into two groups: 1) Near and above $1~GeV$, and 
2) in the region $1.3~GeV \div 1.5~GeV$. The first group scalar mesons form
an $SU(3)$ nonet, which contains two isosinglets, an isotriplet and two
strange isodoublets. In the quark model, the flavor structure of these
scalar mesons would be

\bea
\label{e8201}
\ve \sigma > \es \cos\theta \ve \bar{n}n > - \sin\theta \ve \bar{s}s > ~, \nnb \\
\ve f_0 >    \es \cos\theta \ve \bar{s}s > + \sin\theta \ve \bar{n}n > ~, \nnb \\
 a_0^0 \es \frac{1}{\sqrt{2}} (\bar{u}u - \bar{d}d )~,
~~~a_0^+ = u \bar{d}~,~~~ a_0^- = \bar{d} u~, \nnb \\
\kappa^+ \es \bar{s}u~,~~~\bar{\kappa}^0 = \bar{d}s~,~~~\kappa^- =
\bar{u}s~,
\eea
where $\ve \bar{n}n > = (\bar{u}u + \bar{d}d )/\sqrt{2}$, and $\theta$ is
the mixing angle. Here we take into account the fact that between isoscalars
$\bar{s}s$ and $\bar{u}u+\bar{d}d$ there is mixing, which follows from
experiments. Indeed the observation 
\bea
\Gamma(J/\psi \rar f_0 \omega) \simeq \frac{1}{2} \Gamma(J/\psi \rar f_0
\phi)~,\nnb
\eea
indicates that the quark content of $f_0(980)$ is not purely $\bar{s}s$
state, but should have non--strange parts too \cite{R8202}. 
Secondly, if $f_0(980)$ is purely $\bar{s}s$ state, then $f_0 \rar \pi \pi$ 
should be OZI suppressed.
But the decay width of $f_0(980)$ is dominated by $f_0 \rar \pi \pi$ which
leads to the conclusion that in $f_0(980)$ there should be $\bar{n}n$ parts
as well. Therefore $f_0$ should be a mixture of $\bar{s}s$ and $\bar{n}n$,
as is presented in Eq. (\ref{e8201}). Analysis of the experimental data
shows that the mixing angle $\theta$ lies in the range $25^0 < \theta <
40^0$ and $140^0 < \theta < 165^0$ \cite{R8203}.

Although there is another scenario where mesons below or about $1~GeV$ is 
described as a four--quark state (see for example \cite{R8204}), in this 
work we restrict ourselves to considering the $\bar{q}q$ description
for $f_0(980)$ meson, but taking into account the mixing between
$\bar{s}s$ and $\bar{n}n$. In the present work we study the semileptonic
decays $B^+ \rar f_0(980) \ell^+ \nu$, $D^+_{d,s} \rar f_0(980) 
\ell^+ \nu$ decays in order to get information about the quark content of
$f_0(980)$.

From theoretical point of view, investigation of the semileptonic decays 
is simpler compared to that of hadronic decays, because leptons do not 
participate in strong interactions. The experimental study of weak 
semileptonic decays of heavy flavored mesons is very important for the 
more accurate determination of the Cabibbo-kobayashi-Maskawa (CKM) matrix 
elements, their leptonic decay constants, etc.

The precise determination of the CKM matrix elements depends crucially on
the possibility of controlling long distance interaction effects. So, in
study of the exclusive semileptonic decays the main problem is
calculation of the transition form factors, which involve the long distance
QCD dynamics, belonging to the non--perturbative sector of QCD. For this
reason, in calculation of the transition form factors some kind of
non--perturbative approach is needed. Among all non--perturbative approaches QCD
sum rules method \cite{R8205} is more powerful, since it is based on the
first principles of QCD. About the most recent status of QCD sum rules, the
interested readers are advised to consult \cite{R8206}.

Semileptonic decays $D \rar \bar{K}^0 e \bar{\nu}_e$ \cite{R8207},
$D^+ \rar K (K^{0\ast}) e^+ \nu_e$ \cite{R8208}, $D \rar \pi e \bar{\nu}_e$
\cite{R8209}, $D \rar \rho e \bar{\nu}_e$ \cite{R8210}, $B \rar D(D^\ast)
\ell \bar{\nu}_\ell$ \cite{R8211} and $D \rar \phi \ell \bar{\nu}_\ell$
\cite{R8212} are all studied in the framework of 3--point QCD sum rules 
method.

In this work we study the semileptonic $B_u \rar f_0(980) \ell^+ \nu_\ell$
and $D_{s(d)} \rar f_0(980) \ell^+ \nu_\ell$ decays in the 3--point QCD
sum rules method. The paper is organized as follows: In section 2, we derive 
the sum rules for the form factors, responsible for pseudoscalar to scalar
meson transition. Section 3 is devoted to the numerical analysis of the
transition form factors and discussion and contains our 
conclusions.        

\section{Pseudoscalar--scalar meson transition form factors from 
QCD sum rules} 

Pseudoscalar--scalar transition form factors are defined via the matrix
element of the weak current sandwiched between initial and final meson
states $\lla S(\pp) \vel \bar{q}_1 \gamma_\mu (1-\gamma_5) q_2 \ver
P(p) \rra$, where $q_1$ and $q_2$ are the relevant quarks, $P$ and $S$ are
the pseudoscalar and scalar meson states, respectively. It follows from 
parity conservation in strong interaction that only axial part of weak
current gives non--zero contribution to this matrix element, and imposing
Lorentz invariance, it can be written in terms of the form factors as
follows:
\bea
\label{e8202}
\lla S(\pp) \vel \bar{q}_1 \gamma_\mu (1-\gamma_5) q_2 \ver P(p) \rra =
- i \Big[f_+(p+\pp)_\mu + f_- q_\mu \Big]~,
\eea
where $q_\mu = p_1-p_2$.

For evaluation of these form factors in the QCD sum rule, we consider 
the following the 3--point correlation function
\bea
\label{e8203}
\Pi_\mu (p^2,\pps,q^2) = - \int d^4x \,d^4y \, e^{i(\pp y-px)}
\lla 0 \vel T\Big\{ J_S(y) J_\mu^A(0) J_P(x) \Big\} \ver 0 \rra~,
\eea
where, $J_S=\bar{q}_2q_2$, $J_\mu^A=\bar{q}_2\gamma_\mu \gamma_5 q_1$ and
$J_P=\bar{q}_1 \gamma_5 q_2$ are the interpolating currents of scalar and
pseudoscalar mesons, and weak axial currents, respectively. It should be
noted here that, $q_3=u$, $q_2=u$ and $q_1=b$ for the 
$B_u \rar f_0(980)$ transition; and $q_3=s(d)$, $q_2=s(d)$ and $q_1=c$
for the $D_{s(d)} \rar f_0(980)$ transition, respectively.

The decomposition of the correlation function (\ref{e8203}) into the Lorentz
structures, obviously, has the form
\bea
\label{e8204}
\Pi_\mu = \Pi_+(p+\pp)_\mu + \Pi_-(p-\pp)_\mu~.
\eea

For the amplitudes $\Pi_+$ and $\Pi_-$, we have the following dispersion
relation
\bea
\label{e8205}
\Pi_\pm (p^2,\pps,Q^2) = - \frac{1}{(2\pi)^2} \int 
\frac{\rho_\pm (s,\sp,Q^2) ds \, d\sp}
{(s-p^2)(\sp-\pps)} + \mbox{\rm subtraction terms}~,
\eea 
where $\rho_\pm$ is the corresponding spectral density and $Q^2=-q^2 > 0$.
According to QCD sum rules approach, the correlation function is calculated
by the operator product expansion (OPE) at large Euclidean momenta 
$p^2$ and $\pps$ on one side, and on the other side it is calculated
by inserting a complete set of intermediate states having the same quantum
numbers with the currents $J_S$ and $J_P$.

The phenomenological part of (\ref{e8203}) is obtained by saturating
correlator it with the lowest pseudoscalar (in our case $B_u$, $D_s$ or $D$
mesons) and scalar $f_0(980)$ mesons, yielding
\bea
\label{e8206}
\Pi_\mu = \frac{\lla 0 \vel J_S \ver S(\pp) \rra
\lla S(\pp) \vel J_\mu^A(0) \ver P(p) \rra
\lla P(p) \vel J_P(x) \ver 0 \rra}
{(m_S^2-\pps) (m_P^2-p^2)} +
\mbox{\rm excited states}~.
\eea
The matrix elements in Eq. (\ref{e8206}) are defined as
\bea
\label{e8207}
\lla 0 \vel J_S \ver S(\pp) \rra \es \lambda_S~,\nnb \\
\lla P \vel J_P \ver 0 \rra \es - i \frac{m_P^2 f_P}{m_1+m_2}~,
\eea
where $f_S$ and $f_P$ are the leptonic decay constants of scalar and
pseudoscalar mesons, and $m_S$ and $m_P$ are being their masses, respectively.
Note that, leptonic decay constant $f_S$ in Eq. (\ref{e8207}) is scale
dependent for which we choose the scale to be $\mu=1~GeV^2$, and
\bea
m_1 \es \left\{ \begin{array}{lcl}
m_b & \mbox{\rm for} & B_u \rar f_0 \ell \nu~,\\ \\
m_c & \mbox{\rm for} & D_s \rar f_0 \ell \nu~,~~~
D \rar f_0 \ell \nu~,\end{array} \right.
\nnb \\ \nnb \\ \nnb \\
m_2 \es \left\{ \begin{array}{lcl}
m_u & \mbox{\rm for} & B_u \rar f_0 \ell \nu~,~~~
D \rar f_0 \ell \nu~,\\ \\
m_s & \mbox{\rm for} & D_s \rar f_0 \ell \nu~,
\end{array} \right.\nnb           
\eea

Using Eqs. (\ref{e8202}), (\ref{e8204}), (\ref{e8206}) and (\ref{e8207}),
for the invariant structures we get
\bea
\label{e8208}
\Pi_\pm = - \frac{f_P m_P^2}{m_1+m_2} \,
\frac{\lambda_S f_\pm}{(m_S^2 - \pps) (m_P^2-p^2)}~.
\eea

From QCD side, the correlation function can be calculated with the help of
the OPE at short distance, and in this work we will consider operators
up to dimension six. The theoretical part of the correlator for the
$B_s \rar D_{s_0}(2317) \ell \nu$ is calculated in \cite{R8213}, and
in the present work, for the theoretical part of the corresponding sum 
rules, we will use the results of this work.

For the spectral densities we have
\bea
\label{e8209}
\Pi_+ \es \frac{N_c}{4 \lambda^{1/2}(s,\sp,Q^2)} \Big[\ga\Delta^\prime +
\Delta\dr (1+A+B) + (m_1^2+2 m_1 m_2 + Q^2) (A+B) \Big]~, \\
\label{e8210}
\Pi_- \es \frac{N_c}{4 \lambda^{1/2}(s,\sp,Q^2)} \Big[ \ga \Delta^\prime +
\Delta + m_1^2+2 m_1 m_2 + Q^2 + 2 m_1 m_2 \dr (A-B) \nnb \\
\ar \Delta^\prime  - \Delta - 2 m_1 m_2 \Big]~,
\eea
where $N_c=3$, $\Delta=s-m_1^2$, $\Delta^\prime=\sp - m_2^2$, and
\bea
A \es \frac{1}{\lambda(s,\sp,Q^2)} \Big[ -(s+\sp+Q^2) \Delta^\prime + 2 \sp
\Delta \Big]~, \nnb \\
B \es \frac{1}{\lambda(s,\sp,Q^2)} \Big[ -(s+\sp+Q^2) \Delta^\prime + 2 s   
\Delta^\prime \Big]~. \nnb
\eea
For the decays under consideration, $m_2$ is $m_u(m_d)$ or $m_s$, and
therefore,  to take into account $SU(3)$--violating effects, here and 
in all following calculations we will retain terms that
linear with $m_2$, while neglecting the terms higher order in $m_2$.

For power corrections (PC)we get
\bea
\label{e8211}
\Pi_+^{PC} \es 
\frac{1}{2} \la \bar{q}_2 q_2 \ra \frac{m_1-m_2}{r r^\prime} +
\frac{1}{4} m_2 \la \bar{q}_2 q_2 \ra \Bigg( \frac{m_1^2}{r^2 r^\prime} 
- \frac{2}{r r^\prime} \Bigg) \nnb \\
\ek \frac{1}{12} m_0^2 \bar{q}_2 q_2 \ra \Bigg[\frac{3m_1^2 (m_1-m_2)} 
{r^3 r^\prime} + \frac{2 (m_1-2 m_2)} {r r^{\prime 2}} +
\frac{2 (2 m_1-m_2)} {r^2 r^\prime} \nnb \\
\ar \frac{m_1 (2 m_1^2+m_1 m_2 +2 Q^2) - 2 m_2(m_1^2+Q^2)}
{r^2 r^{\prime 2}} \Bigg] \nnb \\
\ar \frac{4}{81}\pi \alpha_s \la \bar{q}_2 q_2 \ra^2 \Bigg[ 
- \frac{ 12 m_1^3(m_1-m_2)} {r^4 r^\prime}
+\frac{ 8 m_1 m_2(m_1^2+Q^2)} {r^2 r^{\prime 3}}
+\frac{ 56 m_1 m_2} {r r^{\prime 3}} \nnb \\
\ek \frac{4 m_1^2 (2 m_1^2+m_1 m_2 +2 Q^2) - 8 m_1 m_2(m_1^2+Q^2)}        
{r^3 r^{\prime 2}} - \frac{ 8 m_1(8m_1-7m_2)} {r^3 r^\prime} \nnb \\
\ar \frac{48}{r r^{\prime 2}} + \frac{48}{r^2 r^\prime}
- \frac{ 4 (5m_1^2-20 m_1 m_2-2Q^2)} {r^2 r^{\prime 2}} \Bigg] \nnb \\
\ar \frac{1}{9} m_0^2 m_2 \la \bar{q}_2 q_2 \ra^2 \Bigg[
- \frac{ m_1^2(m_1^2+Q^2)} {r^3 r^{\prime 2}} + 
\frac{ 5 m_1^2+4Q^2} {r^2 r^{\prime 2}} +
\frac{ 6 m_1^4} {r^4 r^\prime} +
\frac{ 10 m_1^2} {r^3 r^\prime} \Bigg]~, \\ \nnb \\ \nnb \\
\label{e8212}
\Pi_-^{PC} \es 
- \frac{1}{2} \la \bar{q}_2 q_2 \ra \frac{m_1+m_2}{r r^\prime} +
\frac{1}{4} m_1 m_2 \la \bar{q}_2 q_2 \ra \Bigg( - \frac{m_1}
{r^2 r^\prime}\Bigg) \nnb \\
\ar \frac{1}{12} m_0^2 \bar{q}_2 q_2 \ra \Bigg[\frac{3m_1^2 (m_1+m_2)} 
{r^3 r^\prime} + \frac{2 (m_1+3 m_2)} {r r^{\prime 2}} +
\frac{2 (3 m_1+m_2)} {r^2 r^\prime} \nnb \\
\ar \frac{m_1 (2 m_1^2+m_1 m_2 +2 Q^2) + 2 m_2(m_1^2+Q^2)}
{r^2 r^{\prime 2}} \Bigg] \nnb \\
\ar \frac{1}{81}\pi \alpha_s \la \bar{q}_2 q_2 \ra^2 \Bigg[ 
\frac{ 12 m_1^3(m_1+m_2)} {r^4 r^\prime}
-\frac{ 8 m_1 m_2(m_1^2+Q^2)} {r^2 r^{\prime 3}}
-\frac{ 56 m_1 m_2} {r r^{\prime 3}} \nnb \\
\ar \frac{4 m_1^2 (2 m_1^2+m_1 m_2 +2 Q^2) + 8 m_1 m_2(m_1^2+Q^2)}        
{r^3 r^{\prime 2}} + \frac{ 8 m_1(9m_1+7m_2)} {r^3 r^\prime} \nnb \\
\ar \frac{28 m_1^2}{r^2 r^{\prime 2}} + \frac{8}{r r^{\prime 2}} 
- \frac{8}{r^2 r^\prime} \Bigg] \nnb \\
\ar \frac{1}{9} m_0^2 m_2 \la \bar{q}_2 q_2 \ra^2 \Bigg[
\frac{m_1^2 (m_1^2+Q^2)}{r^3 r^{\prime 2}} -
\frac{m_1^2} {r^2 r^{\prime 2}}
- \frac{6 m_1^4} {r^4 r^\prime}
+ \frac{4}{r r^{\prime 2}} - \frac{4}{r^2 r^\prime}
-\frac{ 24 m_1^2} {r^3 r^\prime} \Bigg]~,
\eea
where $r=p^2-m_1^2$ and $r^\prime=\pps$.
Note that the $D_s \rar f_0(980) \ell^+ \nu_\ell$ and 
$D \rar f_0(980) \ell^+ \nu_\ell$ decays which are considered in
\cite{R8214} differ from our results in three aspects:

\begin{itemize}

\item Our result on spectral density is two times smaller compared to that
given in \cite{R8214}. Since it is known that the main contribution to 
the sum rules comes from the spectral density, it is indispensable that our
results on the form factors differ from those predicted in
\cite{R8214}.

\item In \cite{R8214}, part of those diagrams which are proportional 
to $m_s$ are not taken into account (in our case they correspond to the 
terms proportional to $m_2 m_0^2 \la \bar{q}_2 q_2 \ra$).

\item Sum rules for the form factor $f_-$ are totally absent in \cite{R8214},
which could be essential for the $B_u \rar f_0(980) \tau \nu_\tau$ decay. 

\end{itemize}
  
Contribution of higher states in the physical part of the sum rules are
taken into account with the help of the hadron--quark duality, i.e.,
corresponding spectral density for higher states is equal to the
perturbative spectral density for $s_0$ and $\sp_0$ starting from $s>s_0$
and $\sp>\sp_0$, where $s$ and $\sp$ ar the continuum thresholds in the
corresponding channels.

Equating the two representations for the invariant structures $\Pi_\pm$, and
applying double Borel transformation on the variables $p^2$ and $\pps$ ($p^2
\rar M^2$, $\pps \rar M^{\prime 2}$) in order to suppress the higher states
and continuum contributions, we get the following sum rules for the form
factors $f_+$ and $f_-$:
\bea
\label{e8213}
f_\pm (q^2) \es - \frac{m_1+m_2}{f_P m_P^2} \frac{1}{\lambda_S} e^{m_P^2/M^2} 
e^{m_S^2/M^{\prime 2}} \Bigg\{ \int ds \, d\sp \rho_\pm(s,\sp,Q^2)
e^{-s/M^2 - \sp/M^{\prime 2}} \nnb \\
\ar {\cal B}_{M^2} {\cal B}_{M^{\prime 2}} \Pi_\pm^{PC} \Bigg\}~.
\eea
The double Borel transformation for the quantity $1/r^n r^{\prime m}$ is
defined as:
\bea
\label{e8214}
{\cal B}_{M^2} {\cal B}_{M^{\prime 2}} \frac{1}{r^n r^{\prime m}} =
(-1)^{n+m} \frac{\ga M^2 \dr^{n-1}}{\Gamma(n)}
 \frac{\ga M^{\prime 2} \dr^{m-1}}{\Gamma(m)}
e^{-m_1^2/M^2}~.
\eea

The integration region for the perturbative contribution is determined from
the following inequalities:
\bea
\label{e8215}
-1 \le \frac{2s\sp + (m_1^2-s)(s+\sp+Q^2)}{\lambda^{1/2}(s,\sp,Q^2)
(m_1^2-s)} \le 1~.
\eea

In the calculation of the widths of the considered decays, it is necessary
to know the $q^2$ dependence of the form factors in the whole physical region
$m_\ell^2 \le q^2 \le q_{max}^2$. 

\section{Numerical analysis}
In this section we present our results for the form factors
$f_+(q^2)$ and $f_-(q^2)$ for the decays under consideration.
The main input parameters for the sum rules are the Borel parameters $M^2$
and $M^{\prime 2}$, continuum thresholds $s_0$ and $s_0^\prime$. The values
of other parameters needed are: $m_b=(4.7 \pm 0.1)~GeV$ \cite{R8206},
$m_c=1.4~GeV$, $m_s=0.15~GeV$, $\la \bar{u}u \ra \ve_{\mu=1~GeV} =
-(0.243)^3~GeV^3$, $\la \bar{s}s \ra = 0.8 \times \la \bar{u}u \ra$
\cite{R8215}.
The values of the leptonic decay constants of $B_u$, $D_s$ and $D$ mesons
are determined from the analysis of the corresponding two--point
correlators: $f_{B_u}=(0.14 \pm 0.01~GeV$ \cite{R8216},  
$f_{D_s}=(0.22 \pm 0.02~GeV$ \cite{R8217} and $f_D=(0.17 \pm 0.02~GeV$
\cite{R8206,R8216,R8217}. For the continuum thresholds we take the values
$s_0^{B_u}=(33\pm 2)~GeV^2$, $s_0^{D_s}=(7.7\pm 1.1)~GeV^2$,
$s_0^D=(6\pm 0.2)~GeV^2$ and $\sp_0=1.6~GeV^2$ which is
determined from 2--point sum rules analysis \cite{R8206,R8214,R8216,R8218}.  

The Borel parameters $M^2$ and $M^{\prime 2}$ are the auxiliary parameters
and therefore the physical quantities should be independent of them. For
this reason we need to find the working regions of $M^2$ and $M^{\prime 2}$
where form factors are practically independent of them.

In obtaining the working regions of $M^2$ and $M^{\prime 2}$ the following
two conditions should be satisfied:
\begin{itemize}
\item The continuum contribution should be small, and,
\item power corrections should be convergent.
\end{itemize}
Our numerical analysis shows that, both conditions are satisfied in the
region $10~GeV^2 \le M^2 \le 20~GeV^2$ for $B_u \rar f_0 \ell
\bar{\nu}_\ell$, $4~GeV^2 \le M^2 \le 8~GeV^2$ for $D_s(D) \rar f_0 \ell
\bar{\nu}_\ell$, and $1.2~GeV^2 \le M^{\prime 2} \le 2~GeV^2$ for all
channels.

Varying the input parameters $s_0$, $\sp_0$, $f_0$, $f_{D_s}$, $f_B$ and
$f_D$ in the respective regions as mentioned in the text, we get the
following results for the form factors at $q^2=0$
\bea
\label{e8216}
f_+^{B_u} (0) \es   1.7 (0.25 \pm 0.02)~, \nnb \\
f_-^{B_u} (0) \es - 1.7 (0.24 \pm 0.03)~, \nnb \\
f_+^D (0)     \es   1.7 (0.32 \pm 0.03)~, \nnb \\
f_+^{D_s} (0) \es   1.7 (0.27 \pm 0.02)~,
\eea
The multiplying factor $1.7$ corresponds to the case for
$\lambda_S=0.19~GeV^2$, and without this factor $\lambda_S=0.35~GeV^2$
\cite{R8203}. For a comparison we present the results of the form factor
$f_+$ for the $B_u \rar f_0(980)$ transition coming from the covariant light
front dynamics \cite{R8219} and dispersion relation approach \cite{R8220}, 
as well as the result for the $B_u \rar \pi$ transition \cite{R8221}.
\bea
\label{e8217}
f_+^{B_u\rar f_0}(0) \es 0.27~~~\cite{R8219}~, \nnb \\
f_+^{B_u\rar f_0}(0) \es 0.09~~~\cite{R8220}~, \nnb \\
f_+^{B_u\rar \pi}(0) \es 0.25~~~\cite{R8221}~.
\eea
From a comparison of Eqs. (\ref{e8216}) and (\ref{e8217}) we see that, our
prediction on $f_+$ for the $B_u \rar f_0(980)$ transitions quite close to
the prediction of the light front dynamics and that of $B_u \rar \pi$
transition when $\lambda_S=0.35$, and approximately three times larger
compared to that of the dispersion relation approach. These close results
of the form factor $f_+$ for the $B_u\rar \pi$ and $B_u\rar f_0$
transitions could indicate of the that $\lambda_S$ should have the value
$\lambda_S=0.35$, which is obtained in \cite{R8203} by taking 
${\cal O}(\alpha_s)$ corrections into account.  

Note that we present the form factor $f_-$ only for the $B_u \rar f_0 \tau
\bar{\nu}_\tau$ decay, because this form factor can give considerable
contribution to this decay.

In estimating the width of $P \rar f_0(980) \ell \bar{\nu}_\ell$ decay, we
need to know the $q^2$ dependence of the form factors $f_+(q^2)$ and
$f_-(q^2)$ in the whole kinematical region $m_\ell^2 \le q^2 \le
(m_P-m_{f_0})^2$. The $q^2$ dependence of the form factors can be
calculated from QCD sum rules (see \cite{R8208,R8209}). Unfortunately QCD
sum rule cannot reliably predict $q^2$ dependence of the form factors in the
full kinematical region. The QCD sum rules can reliably predict $q^2$
dependence of the form factors in the
region approximately $1~GeV^2$ below the perturbative cut. In order to
extend the dependence of the form factors on $q^2$ to the full kinematical
region, we look such a parametrization of the form factors where it
coincides with the sum rules prediction of in the above--mentioned region.
Our numerical
calculations shows that the best parametrization of the form factors with
respect to $q^2$ are as follows:
\bea
\label{e8218}
f_P(q^2) = \frac{f_P(0)}{\ds 1 - a_P \hat{q} + b_P \hat{q}^2 - c_P \hat{q}^3
+ d_P \hat{q}^4}~,
\eea
where $P = B_u,~D_s,~D$ and $\hat{q} = q^2/m_P^2$.
The values of the parameters $f_P(0)$, $a_P$, $b_P$, $c_P$ and $d_P$ at
$\lambda_S=0.19~GeV^2$, are given in table 1.   
\begin{table}[h]    
\renewcommand{\arraystretch}{1.5}
\addtolength{\arraycolsep}{3pt}
$$
\begin{array}{|l|c|c|c|c|c|c|}  
\hline
    & f_+(0)             & f_-(0)            & a    & b     & c     & d     \\ \hline
D_s & 1.7\times 0.27     &                   & 0.87 & -0.17 &  0.37 &  1.46 \\ \hline
D   & 1.7\times 0.32     &                   & 0.89 & -0.40 &  0.18 & -1.00 \\ \hline
B_u & 1.7\times 0.25     &                   & 0.48 & -0.30 & -0.47 & -0.99 \\ \hline
B_u &                    & - 1.7\times 0.24  & 0.41 & -0.42 & -0.95 & -1.55 \\ \hline
\end{array}
$$
\caption{Form factors for the $D_s \rar f_0 \ell \bar{\nu}_\ell$,
$D \rar f_0 \ell \bar{\nu}_\ell$ and $B_u \rar f_0 \ell \bar{\nu}_\ell$
decays in a four--parameter fit.}
\renewcommand{\arraystretch}{1}
\addtolength{\arraycolsep}{-3pt}
\end{table}

The dependence of the form factors $f_+$ and $f_-$ (for $B_u \rar f_0
\tau \bar{\nu}_\tau$ decay) are given in Figs. (1)--(4).

Using the parametrization of Eq. (\ref{e8202}), for the 
$P \rar f_0 \ell \bar{\nu}_\ell$ differential decay width, we get   
\bea
\label{e8219}
\frac{d\Gamma}{dq^2} \es \frac{A}{192 \pi^3 m_{P}^3} G^2 \vel V_{ij}
\ver^2 \lambda^{1/2}(m_P^2,m_{f_0}^2,q^2) \ga \frac{q^2-m_\ell^2}{q^2}
\dr^2 \nnb \\
\cp \Bigg\{ -\frac{(2 q^2+m_\ell^2)}{2}  \Big[ \vel f_+(q^2) \ver^2 (2
m_P^2 + 2 m_{f_0}^2 - q^2 ) + 2 (m_P^2 - m_{f_0}^2)
\mbox{\rm Re} [f_+(q^2) f_-^\ast(q^2)] \nnb \\
\ar \vel f_-(q^2) \ver^2 q^2 \Big]
+ \frac{(q^2+ 2 m_\ell^2)}{q^2} \Big[ \vel f_+(q^2) \ver^2 (m_P^2 -
m_{f_0}^2)^2 \nnb \\
\ar 2 (m_P^2 - m_{f_0}^2) q^2 
\mbox{\rm Re} [f_+(q^2) f_-^\ast(q^2)] + \vel f_-(q^2) \ver^2 q^4 \Big]
\Bigg\}~,
\eea
where
\bea
A \es \left\{ \begin{array}{lclcl}
\cos^2\theta & \mbox{\rm for} & D_s \rar f_0 \ell \bar{\nu}_\ell & \mbox{\rm and}~,& \\ \\
\ds\frac{\sin^2\theta}{2} & \mbox{\rm for} & D \rar f_0 \ell \bar{\nu}_\ell &
\mbox{\rm and} & B_u \rar f_0 \ell \bar{\nu}_\ell~,\end{array} \right.
\nnb \\ \nnb \\ \nnb \\
V_{ij} \es \left\{ \begin{array}{lcl}
\vel V_{ub} \ver = 4.31\times10^{-3} & \mbox{\rm for} & B_u \rar f_0 \ell
\bar{\nu}_\ell~,  \\ \\
\vel V_{cs}\ver = 0.96  & \mbox{\rm for} & D_s \rar f_0 \ell \bar{\nu}_\ell~,\\ \\
\vel V_{cd} \ver = 0.23  & \mbox{\rm for} & D   \rar f_0 \ell \bar{\nu}_\ell~, \\  
\end{array} \right.\nnb           
\eea

Taking into account the $q^2$ dependence of the form factors $f_+$ and $f_-$
and performing integration over $q^2$ and using the lifetimes of $B_u$,
$D_s$ and $D$ mesons, we get the following values for the branching ratios
when $\lambda_S = 0.19~GeV^2$:
\bea
\label{e8220}
{\cal B}(B_u \rar f_0 \tau \bar{\nu}_\tau) \es 
\frac{\sin^2\theta}{2}\times(7.17 \times 10^{-5})~,\nnb \\  
{\cal B}(B_u \rar f_0 \mu \bar{\nu}_\mu) \es 
\frac{\sin^2\theta}{2}\times(2.1 \times 10^{-4})~,\nnb \\
{\cal B}(B_u \rar f_0 e \bar{\nu}_e) \es 
\frac{\sin^2\theta}{2}\times(2.1 \times 10^{-4})~,\nnb \\
{\cal B}(D_s \rar f_0 \mu \bar{\nu}_\mu) \es 
\cos^2\theta\times(3.85 \times 10^{-3})~,\nnb \\
{\cal B}(D_s \rar f_0 e \bar{\nu}_e) \es 
\cos^2\theta\times(4.07 \times 10^{-3})~,\nnb \\
{\cal B}(D \rar f_0 \mu \bar{\nu}_\mu) \es 
\frac{\sin^2\theta}{2}\times(4.98 \times 10^{-4})~,\nnb \\
{\cal B}(D \rar f_0 e \bar{\nu}_e) \es 
\frac{\sin^2\theta}{2}\times(5.29 \times 10^{-4})~.
\eea

We see from (\ref{e8220}) that the ratios of the widths
\bea
R_1 \es \frac{{\cal B}(D   \rar f_0 \ell \bar{\nu}_\ell)}
             {{\cal B}(D_s \rar f_0 \ell \bar{\nu}_\ell)}~, \nnb \\
R_2 \es \frac{{\cal B}(B_u \rar f_0 \ell \bar{\nu}_\ell)}
             {{\cal B}(D_s \rar f_0 \ell \bar{\nu}_\ell)}~, \nnb
\eea
are directly related with the mixing angle $\theta$. On the other hand,
as far as the flavor structure of $f_0(980)$, as is given in Eq. (\ref{e8201}), is
considered, the ratio
\bea
R_3 \es \frac{{\cal B}(B_u \rar f_0 \ell \bar{\nu}_\ell)} 
             {{\cal B}(D   \rar f_0 \ell \bar{\nu}_\ell)}~, \nnb
\eea  
is independent of the mixing angle $\theta$. Therefore, experimental
measurement of the branching ratios of $B_u \rar f_0 \ell \bar{\nu}_\ell$,
$D_s \rar f_0 \ell \bar{\nu}_\ell$ and $D   \rar f_0 \ell \bar{\nu}_\ell$
decays can give direct information about the mixing angle $\theta$, as well
as, about the flavor structure of $f_0(980)$ meson.

In conclusion, we study the semileptonic decay of pseudoscalar mesons to the
scalar $f_0(980)$ meson. The transition form factors are calculated using
3--point QCD sum rule analysis and then we estimate the corresponding
branching ratios.

\newpage

\section*{Figure captions}
{\bf Fig. (1)} The dependence of the form factor $f_+$ on $q^2$ at
$M^2=15~GeV^2$, $M^{\prime 2}=2~GeV^2$, $m_0^2=0.8~GeV^2$, 
$s_0=33~GeV^2$ and $s_0^\prime =
1.6~GeV^2$, for the $B_u \rar f_0(980) \ell \bar{\nu}_\ell$ decay. \\ \\
{\bf Fig. (2)} The dependence of the form factor $f_-$ on $q^2$ at
$M^2=15~GeV^2$, $M^{\prime 2}=2~GeV^2$, $m_0^2=0.8~GeV^2$,
$s_0=33~GeV^2$ and $s_0^\prime =
1.6~GeV^2$, for the $B_u \rar f_0(980) \ell \bar{\nu}_\ell$ decay. \\ \\
{\bf Fig. (3)} The dependence of the form factor $f_+$ on $q^2$ at
$M^2=6~GeV^2$, $M^{\prime 2}=2~GeV^2$, $m_0^2=0.6~GeV^2$,
$s_0=7~GeV^2$ and $s_0^\prime =
1.6~GeV^2$, for the $D_s \rar f_0(980) \ell \bar{\nu}_\ell$ decay. \\ \\
{\bf Fig. (4)} The dependence of the form factor $f_+$ on $q^2$ at
$M^2=6~GeV^2$, $M^{\prime 2}=2~GeV^2$, $m_0^2=0.8~GeV^2$,
$s_0=6~GeV^2$ and $s_0^\prime =
1.6~GeV^2$, for the $D \rar f_0(980) \ell \bar{\nu}_\ell$ decay.

\newpage

\begin{figure}
\vskip 3. cm
    \includegraphics{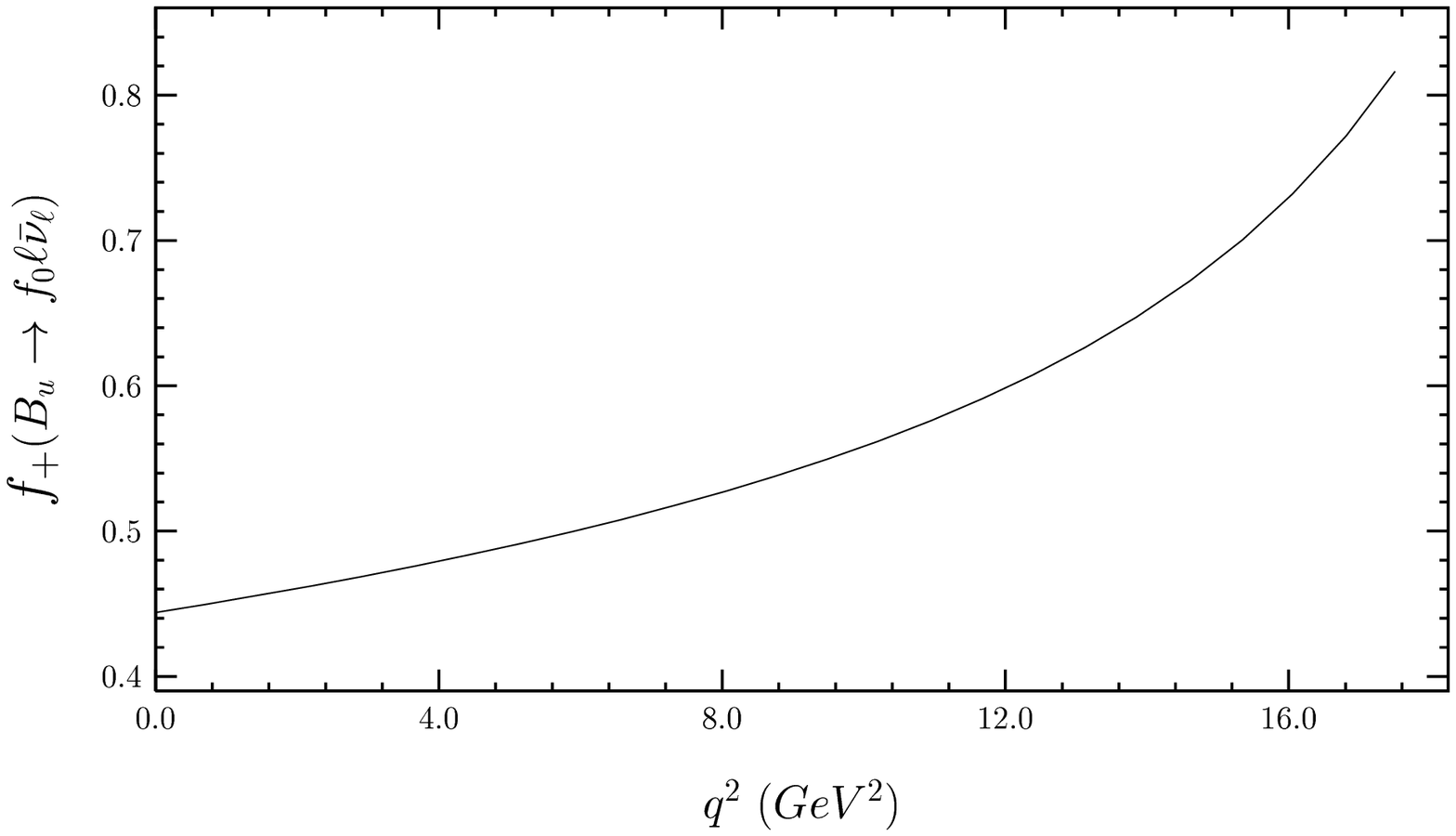}
\vskip 6.3cm
\caption{}
\end{figure}

\begin{figure}
\vskip 4.0 cm
    \includegraphics{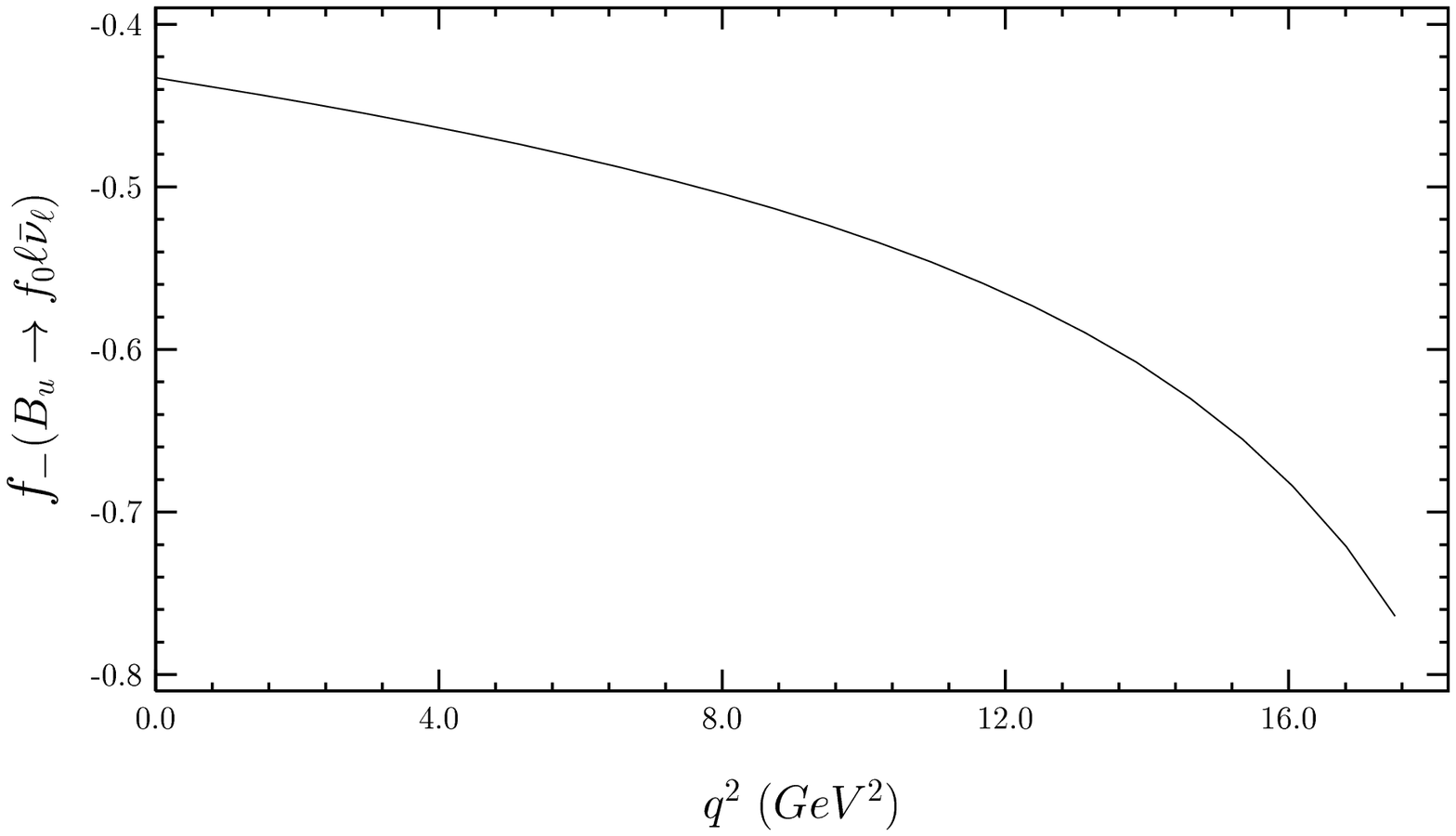}
\vskip 6.3 cm
\caption{}
\end{figure}

\begin{figure}
\vskip 3. cm
    \includegraphics{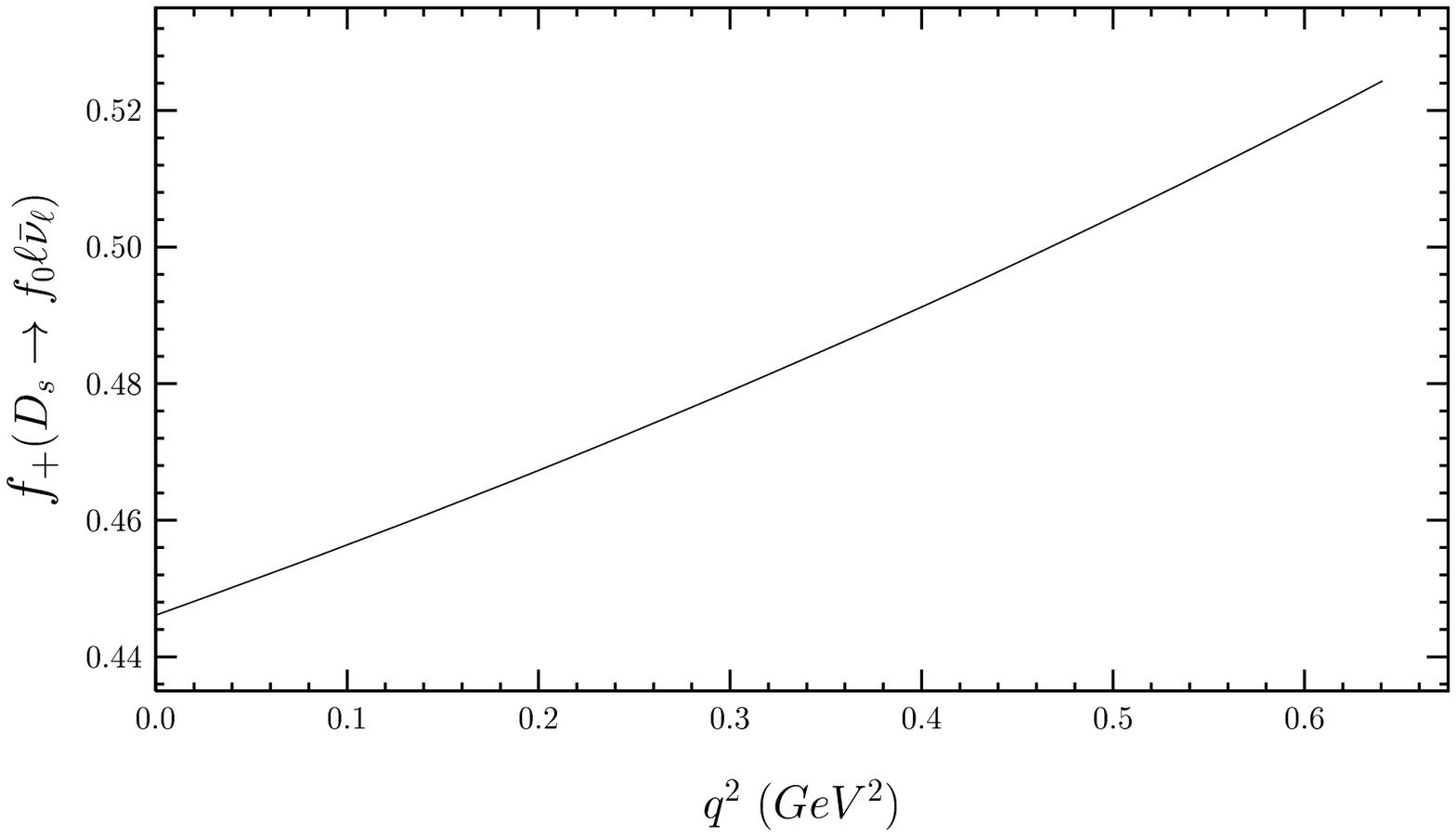}
\vskip 6.3cm
\caption{}
\end{figure}

\begin{figure}
\vskip 4.0 cm
    \includegraphics{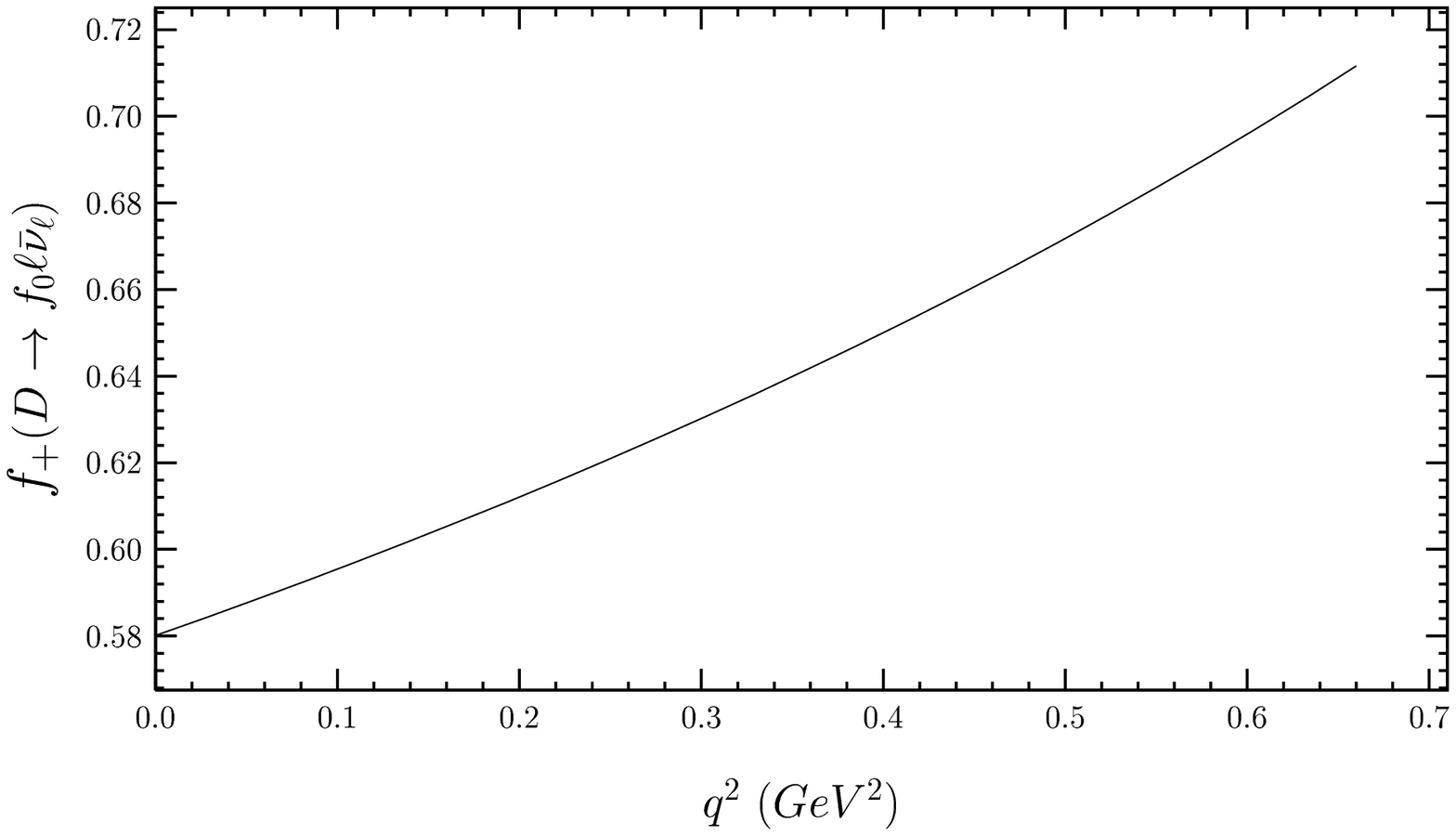}
\vskip 6.3 cm
\caption{}
\end{figure}

\end{document}